\documentclass[superscriptaddress,english,aps,prl,twocolumn,nopacs,groupedaddress,floatfix]{revtex4}

\usepackage{amsmath} 
\usepackage{bbm}
\usepackage{epsfig}
\usepackage{times}
\usepackage{color}
\usepackage{amsfonts}
\usepackage{babel}
\makeatletter

\newcommand{\id}{\mathbbm{1}}
\newcommand{\cc}{{\mathbbm{C}}}

\newcommand{\tr}{\text{tr}}

\graphicspath{{images/}}

\begin{document}

\title{Unitary circuits for strongly correlated fermions}

\newcommand{\unam}{Universidad Nacional Aut\'onoma de M\'exico, M\'exico, D.F., M\'exico}
\newcommand{\ifunam}{\affiliation{Instituto de F\'{\i}sica, \unam}}
\newcommand{\ImperialBell}{\affiliation{Blackett Laboratory, Imperial College London, 
London SW7 2BW, UK}}
\newcommand{\WiKo}{\affiliation{Institute for Advanced Study Berlin, D-14193 Berlin, Germany}}
\newcommand{\uPotsdam}{\affiliation{Institut f\"ur Physik und Astronomie, 
University of Potsdam, D-14476 Potsdam, Germany}}
\author{Carlos Pineda} \uPotsdam \ifunam
\author{Thomas Barthel}\uPotsdam 
\author{Jens Eisert} \uPotsdam \ImperialBell\WiKo
 
\begin{abstract}
We introduce a scheme for efficiently describing pure states of strongly
correlated fermions in higher dimensions using unitary circuits featuring a causal cone. 
A local way of computing local expectation values is presented. We formulate a
dynamical reordering scheme, corresponding to time-adaptive Jordan-Wigner
transformation, that avoids nonlocal string operators. Primitives of such a reordering scheme are highlighted. 
Fermionic unitary circuits can be contracted with the same complexity as in the spin case. 
The scheme gives rise to a variational description of fermionic models not suffering from a
sign problem. We present numerical examples on $9\times 9$ and $6\times 6$ fermionic lattice model to show the
functioning of the approach.
\end{abstract}

\maketitle

Strongly correlated quantum lattice models still pose some of the most
intriguing problems in quantum physics, presumably being at the basis of
phenomena such as high-temperature superconductivity~\cite{Vojta}.  To
describe such models theoretically is often very difficult, an obvious obstacle
being the prohibitive dimension of Hilbert space even for moderately sized
systems when naively representing ground states. In recent years, it has
increasingly become clear, however, that typical ground states of physically
meaningful local models are lurking in some corner of Hilbert space, one that
can often even be identified \cite{Area, Scholl, White, PEPS1, PEPS2, MERA1,
Flow, Rizzi, MERAFreeFermions, 2D, Homogeneous}.  Hence to faithfully describe
such a state, even
though it may be impossible to parametrize the entire physical space, one
merely has to parametrize those quantum states from the relevant set,
requiring significantly fewer parameters.  The most prominent and, to date, most
powerful incarnation of this idea is provided by the density-matrix
renormalization group (DMRG) method  \cite{White,Scholl}, requiring only
linearly many parameters in the system size, but still giving an exceptionally
good description of gapped one-dimensional spin chains and a very reasonable
one for critical chains.

To realize such an idea in higher-dimensional systems is significantly more
difficult, although progress has been made in recent years
\cite{PEPS1,PEPS2,Rage,MERA1,Flow,Rizzi,MERAFreeFermions,2D,Homogeneous,Area,Scholl},
specifically when it comes to generalizing DMRG ideas to higher dimensions,
including variations of tensor product states or projected entangled pair
states, variants with graph enhancement, or the promising multiscale entanglement renormalization (MERA). This
is a scale-invariant approach related to renormalization that has been proposed
in Ref.\ \cite{MERA1} and implemented in Refs.\
\cite{Flow,Rizzi,MERAFreeFermions,Homogeneous,2D}.  Area laws \cite{Area} often
point to the part of Hilbert space that is being occupied. 
 
To describe higher dimensional fermionic systems, such as the Fermi-Hubbard model itself
\cite{Vojta}, in such a fashion, appears particularly promising, but also
particularly challenging.  Here, other key methods of describing quantum
lattice models like the powerful quantum Monte Carlo method  are
hampered by the sign problem \cite{QMC}.  Surely, fermionic models
can readily be represented as spin models, yet at the expense of losing
locality (or by increasing the locality
region of Hamiltonians \cite{Mapping}).  If one considers the  term $f_j^\dagger
f_k$, $j<k$, then its spin representation under the Jordan-Wigner
transformation (JWT) is 
 \begin{equation*}
 	\sigma_j^- \otimes
 	\bigotimes_{j<l<k} \sigma_l^z
 	 \otimes
 	\sigma^+_{k}.
 \end{equation*}
For a pair of nearest-neighbor sites $\langle j,k\rangle$ on a 
$d\times d$ lattice, the occurring string operator that is supported
not only on the spins associated with $j$ and $k$, but in fact on all
spins between $j$ and $k$, will typically have a length linear in $d$.
It should be clear that no order can be chosen to let this apparent problem disappear.

In this work, we present a method for studying strongly correlated fermionic
models using quantum circuits,  that is, circuits of fermionic gates, in a way
that is not overburdened by string operators: When describing the system and
computing local expectation values, one has to deal with operators having
support identical to that of a corresponding spin system (fermionic gates of
the circuit replaced by regular ones), and strings can be made to disappear.
The key point is to acknowledge that while any fixed order will give rise to
the aforementioned problem, we are not necessarily forced to pick any order in
the first place.  Instead, one can employ a dynamical reordering of the
fermionic modes in the essential part of the lattice, reordering and projecting
out particles ``on the fly,'' in dynamical JWTs. We show that this can be
consistently done, not altering expectation values of parity-preserving
operators.  In this way, we find that to describe fermionic lattice models is,
in this sense, just as hard or easy as describing spin models. More formally, the
contraction complexity of the circuit is the same.

{\it Fermions.} We first prepare the ground of the dynamical
reordering idea. The algebra ${\cal G}(L)$ of a set of $n$ fermionic
modes is spanned by products of anticommuting
fermionic operators 
$\{f_j,f_j^\dagger:j=1,\dots,
n\}$, with
$\{f_j,f_k\}=0$, $\{f_j,f_k^\dagger\}=\delta_{j,k}$.
Physical operators form the subalgebra ${\cal F}(L)$, the so-called 
physical algebra, of operators respecting the fermion number parity
\cite{TerhalKitaev,Fermions}. In 
practical terms, this means that they are even polynomials in the 
fermionic creation and annihilation operators.  Hence, the physical 
algebra splits into a direct sum of an even and an odd part,
${\cal F}(L)={\cal F}^\text{(even)}(L)\oplus 
	{\cal F}^\text{(odd)}(L)$.
For a subset $I\subset L$ of some sites, one similarly finds the 
physical algebra ${\cal F}(I)$. 

{\it Fermionic unitary circuits and multiscale entanglement renormalization.} A
fermionic unitary is a parity-preserving unitary gate acting on fermionic
modes, $U=\exp(i H)$, where $H\in {\cal F}(L)$ is a hermitian operator.  In a
circuit \cite{MERA1,Flow,Rizzi,MERAFreeFermions}, unitaries will typically not
have support on the entire lattice, but will be local; that is, it will have a
small support $I$ independent of $n$ such that $U\in {\cal F}(I)$.  A fermionic
unitary circuit is an ordered product of fermionic unitaries.  For most of
this work, we in fact consider conjugation, in particular for the evaluation of
expectation values of local observables $A\in {\cal F}(L)$, with respect to
states that have been prepared by applying a fermionic unitary circuit to the
vacuum $|\o\rangle $,
\begin{equation}\label{E}
	E= \langle \o |
	\dots U_2 U_1 A U_1^\dagger  U_2^\dagger \dots |\o\rangle,
\end{equation}	 
or by applying it to some other with odd fermion number parity, for example,
$f^\dagger_n|\o\rangle$  instead of $|\o\rangle $. In fact, there is a natural discrete time label $t$
in a circuit, in that $t=0,\dots, T$ labels the time at which a single gate is
being applied. What is more, we focus on
circuits that feature a causal cone (see Fig.\ \ref{Fig1}),
most prominently present in the
MERA in the original sense of Ref.\ \cite{MERA1}: The cone is 
formed by those unitaries in a circuit that cannot be sequentially 
canceled with their conjugates from the dual vector in 
Eq.\ (\ref{E}) due to the presence of a local observable $A$ that is supported 
only on a few sites, typically nearest neighbors.
Efficient schemes will have a causal cone of a fixed width, in that a 
local operator will only touch a constant number of unitaries for each time step. The
sequence of sequentially computing the expectation values respecting the discrete 
time order gives rise to a contraction. We 
can also allow for tensors different from unitary gates and partial traces. 
We also allow time steps where partial projections are applied 
onto the fermionic vacuum in some mode. 

We now turn to aspects that are specific for fermions. A first key property from the 
elements of disjoint subalgebras $U \in {\cal F}(I)$ and $V \in {\cal F}(J)$, $I\cap J=\emptyset$, 
respecting parity, is 
that their elements 
commute, 
$[U,V]=0$.
This has the important consequence that just as for spin systems, all
unitaries outside a causal cone can be canceled, and the fermionic variant
inherits the same cone as the spin system. 
We now see that the computation of
the expectation value Eq.\ (\ref{E}) can be done without having to deal with
string operators outside the cone.
 
{\it Jordan-Wigner transformations.} 
Fermionic operators are encoded in an occupation number representation,
a representation necessarily depending on the chosen
order of the fermionic modes. As we will later make updates with different
fermionic orderings at different times, it makes sense to first highlight the
status of the JWT.  For the entire system $L$ one can take some order $O$
corresponding to an element of the symmetric group $S_n$, for example the
trivial order $O=(1,\dots, n)$. For this order $O$, a basis for the Hilbert
space is given in the occupation number representation, 
	$|i_1, \dots, i_n\rangle_O =  
	(f_{O_1}^\dagger)^{i_1}\dots
	(f_{O_n}^\dagger)^{i_n}|\o\rangle$,
giving an identification of the total Hilbert space with the $(\cc^2)^{\otimes
n}$, the Hilbert space of $n$ spins $1/2$. Expressing fermionic operators in
the occupation number representation amounts to a map $J_O: {\cal
G}(L)\rightarrow M(L)$, where $M(L)$ is the algebra of linear operators on
$(\cc^2)^{\otimes n}$. This map is the well-known JWT, depending on $L$ and the chosen order $O\in S_n$
of the fermionic modes; for example,
\begin{equation}
	J_O(f_{O_k}) =
	\biggl(  \bigotimes_{l=1}^{k-1} \sigma_{l}^z \biggr)
	\otimes \sigma_k^+, \\
\end{equation}
where $\sigma_k^{x,y,z}$ denote the familiar Pauli operators and $\sigma_j^+=
(\sigma_j^x + i \sigma_j^y)/2$. In this light, a JWT simply gives fermionic
operators in the occupation number (or spin) representation for a given
ordering $O$.
We hence identify local fermionic operators with non-local
operators in the respective spin systems.  Also, physical operators contain
strings that depend on the order chosen. Consider a local physical 
operator acting on
sites $I$, the spin representation will in general also be supported on the
spins in between with respect to order $O$. 
For example, for the local fermionic number operator one finds
$J_O(f_{O_j}^\dagger f_{O_j})= \sigma_j^z$, but the occupation number
representation of $f_{O_j}^\dagger f_{O_k}$
\begin{equation}
	J_{O}( f_{O_j}^\dagger
	f_{O_k} ) = \sigma_j^- \otimes
	\bigotimes_{j<l<k} \sigma_l^z
	 \otimes
	\sigma^+_{k} 
\end{equation}
with $j<k$, contains a string of Pauli operators on $[j,k]$.

\begin{figure}
  \centering
  \includegraphics[scale=.3]{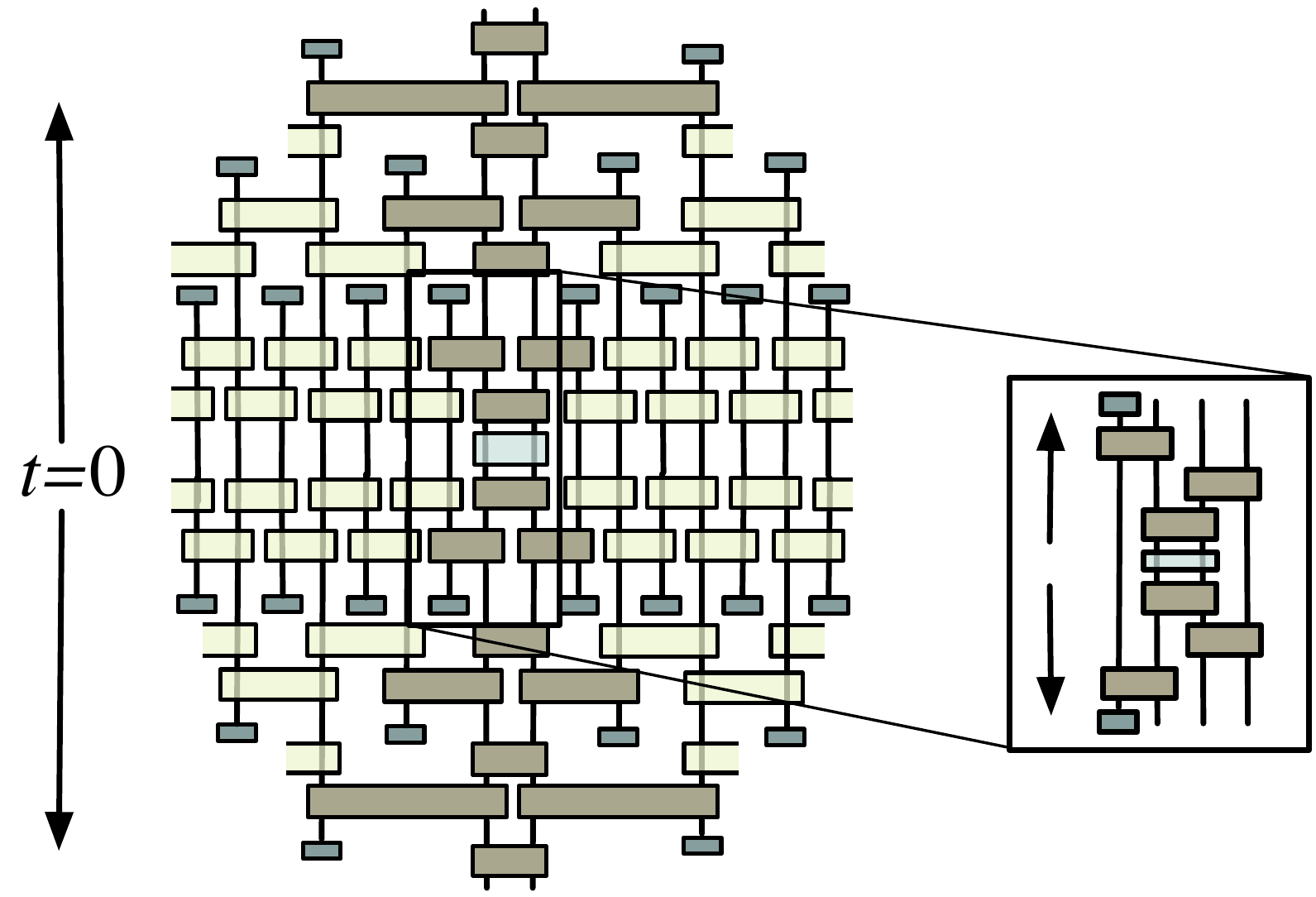}\\
  \caption{
    \label{Fig1}
    Example for the evaluation of a local observable (black box)
    with respect to a fermionic MERA (gray boxes). Each gray box
    represents a single fermionic unitary. Dark gray boxes make up 
    the causal cone of the observable; all others lie outside of it and cancel
    each other. Only for conceptual clarity, a 1D MERA is depicted, not a
    2D one, for which dynamical reordering becomes relevant.}
\end{figure}

For operators $A\in {\cal F}(L)$ parity conservation 
manifests itself in the occupation number representation for an $O$
as follows: Splitting the occupation number basis
into two parts, depending on the parity of $\sum_k i_k$, the spin representation
$a=J_O(A)$ of $A$ attains the form 
 $a =  a^\text{(even)}  \oplus a^\text{(odd)}$.

{\it Dynamical reordering.} We now describe a new scheme to contract a
fermionic unitary circuit, henceforth called dynamical reordering, and
demonstrate that the contraction can be done with essentially the same
complexity compared to the corresponding spin circuit: 
We take the ``time order'' in contracting, with $t=0$
corresponding to the local operator $A$ itself. In turn, in our fermionic 
variant of the MERA \cite{MERA1}, the top layer with largest $t$
defines the parity, similar to fixing the topological degrees of freedom in Kitaev's toric code \cite{Aguado}. 
The general idea of dynamical reordering is
to order only the modes we are currently using at a time $t$ in a nontrivial
fashion with order $O^t$ explicitly dependent on $t$.  As we proceed, new modes
appear in the causal cone, and they are added to the description. When they are
discarded due to projection, they are removed from the description
$O^t$ as $t$ evolves.  Hence, we arrive at an entirely local description of
the fermionic tensor network, using the subsequent rules (see Fig.~\ref{Fig2}).

{\it (a) Local representation of fermionic local operators:} 
Consider a local operator $A\in {\cal F} (I)$; that is, it
is an even polynomial in the operators $\{f_j,f_j^\dagger\}$ for $j\in I$.
The spin representation of the fermionic operator $A$  
in order $O$ is given by 
$a=J_{O}(A)$.  This representation
only involves $k=|I|$ spins.

{\it (b) Reordering fermionic modes:} 
Consider some fermionic operator $A$
and an order $O^t$ at time $t$. 
The new spin representation at a time $t+1$ in the new order
$O^{t+1}$ reads 
$J_{O^{t+1}}(A)= p J_{O^{t}}(A) p$,
with
$p=\prod_i \id \otimes s_{O^t_i, O^t_{i+1}}\otimes \id $,
$s_{O^t_i, O^t_{i+1}}= |0,0\rangle\langle 0,0|
	+|0,1\rangle\langle 1,0|
	+ |1,0\rangle\langle 0,1|
	-|1,1\rangle\langle 1,1|$,
where the $i$ are chosen such that the corresponding sequence of 
nearest-neighbor pair permutations of modes gives rise to the overall
permutation $O^{t+1}= \pi(O^{t})$; so familiar fermionic
swaps can be applied to the local spin representation. 

{\it (c) Prepending of fermionic modes:} 
Let us have an operator $A$ at some time $t$, 
represented as $a=J_{O^{t}}(A)$ in the order $O^{t}$. We now just prepend
a new fermionic mode, yielding the new order $O^{t+1}= (k, O^{t})$. Then the 
new representation of $a'=J_{O^{t+1}}(A)$ is, in the even and odd sectors,
given by
\begin{eqnarray}
(a')^\text{(even or odd)}=
|0\rangle\langle 0| \otimes a^\text{(even or odd)}+
|1\rangle\langle 1| \otimes a^\text{(odd or even)}.\label{ae}
\end{eqnarray}

{\it (d) Conjugation with fermionic unitaries having the same support:}
Consider a subset of modes $I$ at time $t$.  Let $U, A\in {\cal F}(I)$ 
be fermionic operators with support in $I$.
Then
$J_{O^{t+1}}(UAU^\dagger) = J_{O^{t}}(U) J_{O^{t}}(A)  J_{O^{t}}(U^\dagger) $
if the same order $O^{t+1}=O^{t}$ is taken. If $U$ is stored in a different 
order, the permutation rule has to be applied first. 

{\it (e) Partial trace over the first mode:} Let us have some operator $A$ at
time $t$ in order $O^{t}$, and we trace out the first mode $O^{t}_1$; that is, 
$O^t=(O^t_1,O^{t+1})$. Then the spin representation of the resulting fermionic
operator is, with $k=|O^{t}|$,
\begin{eqnarray}
	J_{O^{t+1}}(\tr_{O^{t}_1} A) \nonumber
&=&\sum\langle  i ,j_2, \dots, j_k | J_{O^t}(A)|i  ,i_2 ,\dots ,i_k \rangle\\
&\times& |j_2 ,\dots, j_k \rangle\langle i_2 ,\dots, i_k |.\label{eq:partialTraceSpin}
\end{eqnarray}

{\it (f) Partial projection over the first mode:} 
Let us consider the same orderings $O^t$ and
$O^{t+1}$ as for rule (e), but instead of tracing out mode $r=O^t_1$ from operator $A$, we
now want the projection of the mode to the empty state or that of
a filled mode. The corresponding projectors in ${\cal F}(I)$ are
$P^{(0)}_r=f_r f_r^\dag$ and $P^{(1)}_r=f_r^\dag f_r $, respectively, so that the
projected fermionic operators are $P^{(i)}_rAP^{(i)}_r$, $i=0,1$.
Applying this to Eq.\ \eqref{eq:partialTraceSpin}, we find
\begin{multline}\label{eq:partialProjectSpin}
	J_{O^{t+1}}(\tr_r P^{(i)}_r A P^{(i)}_r) 
=\sum
|j_2,\dots, j_k \rangle\langle i_2,\dots, i_k | \\
\times \langle  i , j_2,\dots, j_k | J_{O^t}(A)|i , i_2,\dots, i_k \rangle.
\end{multline}

\begin{figure}
  \centering
  \includegraphics[scale=.3]{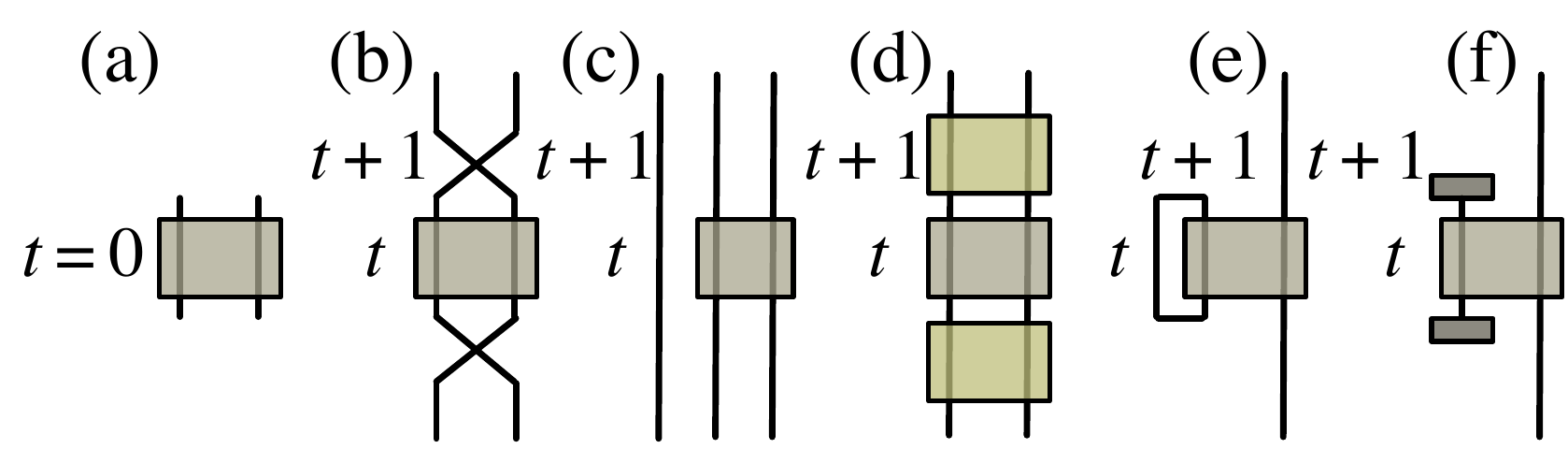}\\
  \caption{
    \label{Fig2}
		The basic primitives of (a) representing
		local operators, (b) reordering fermionic modes, (c) adding fermionic modes, 
		(d) conjugating with unitaries, (e) doing partial traces, and (f) generating partial 
		projections. 
	}
\end{figure}

The main point now is that if one follows this recipe of rules (a)-(f),
and keeps only local
descriptions at each time, involving the nontrivial 
support, one gets a value identical
to Eq.\ (\ref{E}) as if one had performed a global JWT 
and computed it--then with a significant overhead in
complexity.  
To prove the reordering rule for two modes with labels
$j,k$ with $j=O^t_i$ and  $k=O^t_{i+1}$, that is, two modes 
neighboring in the order $O^t$, we need to show
$J_{O^t} (S_{j,k}A S^\dagger_{j,k}) = s_{j,k} J_{O^t} (A ) s^\dagger_{j,k} 
	=  J_{O^{t+1}} (A )$, 
where $S_{j,k}$ acts as $S_{j,k} f_j S^\dagger_{j,k}= f_k$ and $S_{j,k} f_k S^\dagger_{j,k}= f_j$. One 
finds
$S_{j,k}=\id-f_j^\dagger f_j -f_k^\dagger f_k +f_j^\dagger f_k + f_k^\dagger
	f_j$ \cite{TerhalKitaev,footnote}.
The reordering rule for more than two modes follows by iteration.
To derive the rule of adding modes, note that the local JWT gives
$J_{O^t}(f_{O_k^t})= (\otimes_{l=1}^{k-1}\sigma^z_l) \otimes \sigma_k^+$, and
hence we obtain for $O^{t+1}=(x,O^t)$ that 
$J_{O^{t+1}}(f_{O^{t+1}_{k+1}})=
J_{O^{t+1}}(f_{O^t_k})= \sigma_1^z \otimes J_{O^t}(f_{O^t_k})$. 
Exploiting the conservation of the fermion number parity for any $A\in
{\cal F}(I)$, 
$J_{O^{t+1}}(A)=
\sum_\nu (\sigma_1^z)^{2\nu} \otimes
J_{O^t}(A_\nu)= \id\otimes J_{O^t}(A)$, 
where $A_\nu$ collects all monomial terms of degree $2\nu$ such that
$A=\sum_\nu A_\nu$. With $a'=  J_{O^{t+1}}(A)$ and $a= J_{O^{t}}(A)$, for the
even and odd sectors, this gives Eqs.\ (\ref{ae}). To show Eq.\ \eqref{eq:partialTraceSpin},
note that for any operator $B$ not acting on mode $O_1$, that is, $B \in
{\cal F}(I\setminus \{ O_1 \})$, matrix elements obey
$\langle  i_1,\dots ,i_k |_O B|j_1, \dots ,j_k \rangle_O
= \delta_{i_1,j_1}\, 
\langle  i_2,\dots ,i_k |_{O'} 
B|j_2,\dots,  j_k \rangle_{O'}$, where $O=(O_1,O')$ and $k=|O|$. 
With 
\begin{multline}\label{eq:partialTraceFermi}
	\tr_{O_1} (A) =\langle  j_1,j_2\dots,j_k |_O A
|j_1 ,i_2,\dots ,i_k \rangle_{O}\\
\times |j_2 ,\dots, j_k \rangle_{O'}\langle i_2, \dots, i_k |_{O'},
\end{multline}
it follows that $\tr (A B) = \tr[ \tr_{O_1} (A) B]$ for any $A\in{\cal F}(I)$.
This means that 
$\tr_{O_1}$ is the fermionic partial trace for mode $O_1$.  The resulting
occupation number representation $J_{O'}(\tr_{O_1} A)$ of the partial trace is
given in Eq.\ \eqref{eq:partialTraceSpin}.  
Rules (c), (e), and (f) were given for the case where
only the very first mode $O_1$ of the ordering is affected.  
Combining those rules with the rule for reordering
modes, they are generalized to the case where an arbitrary mode $O_j$
is affected. In the corresponding expressions 
this results in extra sign factors.  For the generalization of
rule (d), note that when one wants to perform a
contraction in which the support is not the same for the two involved
operators, one can add the corresponding modes in both operators, reorder and
apply (d). With theses basic building blocks, we can thus contract
fermionic tensor networks and also build, for example, reduced density matrices~\cite{Rizzi}.

{\it Observation (Computing expectation values).}
When computing
expectation values 
 $E= \langle \Psi |
	\dots U_2 U_1 H U_1^\dagger  U_2^\dagger \dots |\Psi\rangle$
where $U=U_1 U_2\dots $ 
is a fermionic circuit on an elementary state vector $|\Psi\rangle$, 
then the support of the causal cone
at each time step is exactly the same as if one had a spin system
with the same topology. Hence, a fermionic circuit acting on a
higher-dimensional fermionic system can be described entirely 
locally with the same time and memory complexity.

This observation renders known algorithms useful even in the case of strongly
correlated fermions, with little modification. Notably, for MERA, the
computational effort  for the computation of a local expectation value scales
as $O(\log l)$ in an $l^D$ cubic lattice and polynomially in the refinement
parameter \cite{MERA1}, hence giving rise to an indeed efficient algorithm, up
to, at most, a small constant the same as for bosons or spins. It is noteworthy
that the entire machinery developed for spin systems can, with slight
modification, be used here.  The preceeding observation is independent of
whether the system is free or strongly correlated.  As it turns out, general
fermionic gates also can be taken account, by directly contracting
appropriately ordered fermionic operators.  One can show that the computational
requirements remain unchanged giving rise to a general theory of fermionic
tensor networks.

\begin{figure}
 \centering
 \includegraphics[scale=.8]{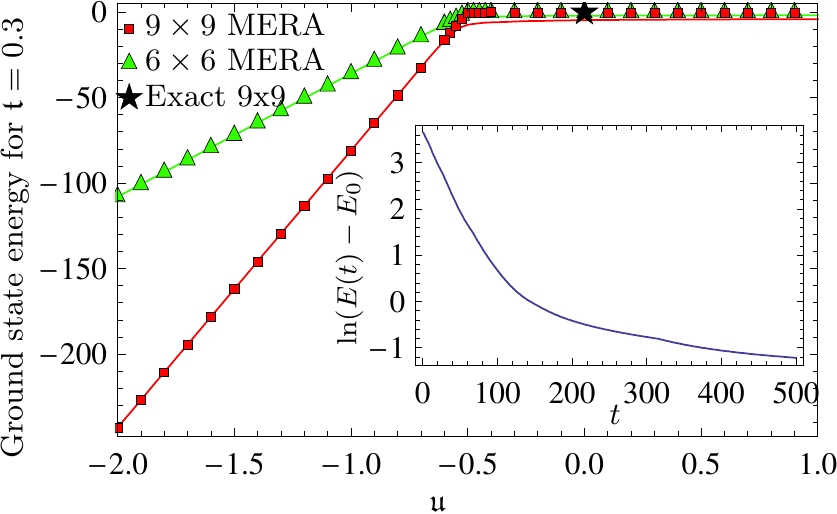}\\
 \caption{ \label{Fig3}
	Numerical examples of the ground-state energy $E_0$ for several
$9\times 9$ and $6\times6$ (boxes and triangles) instances of strongly
correlated fermionic models, $H= {\mathfrak t}\sum_{\langle j,k\rangle}
(f_j^\dagger f_k+h.c.) + \sum_j f_j^\dagger f_j+ {\mathfrak u} \sum_{\langle
j,k\rangle} f_j^\dagger f_j f_k^\dagger f_k$; the free models ${\mathfrak u}=0$
are marked. As the MERA ansatz is a truly variational ansatz, providing upper
bounds to the energy, success is guaranteed by giving lower bounds as 
provided by Anderson bounds (lines).  This is based on an exact diagonalization
of $5\times 5$ fermionic  models, taking care of arising string operators. Even
for this lowest order 2D MERA, in which, in each renormalization step, nine
fermionic modes are mapped to one, we find an impressive precision. The inset
shows the convergence in time for a $9\times 9$ free model ($ {\mathfrak u}=0,
\,{\mathfrak t}=0.3$) in a minimization using imaginary time evolution. In the
same way, correlators and on-site properties can be efficiently computed.
}
\end{figure}

{\it Time evolution.}  One also finds the following: For 
the time evolution, both real and imaginary,
of local fermionic Hamiltonians, we aim at determining
updates of each the 
fermionic unitaries forming the
circuit such that $(U_1') ^\dagger  (U_2') ^\dagger \dots |\psi\rangle$ is 
an optimal approximation to the evolved $\exp(x H) 
U_1^\dagger  U_2^\dagger \dots |\psi\rangle$, again
with the same 
complexity as for a qubit system. 

{\it Numerical implementation.} We have realized a scalable 
numerical implementation of this idea, applied to two-dimensional fermionic 
lattice models. To show the functioning of this approach as a proof of principle, we present benchmark
examples in Fig.\ \ref{Fig3}.

{\it Summary.} In this work, we have introduced a notion of time-adaptive
JWTs allowing us to contract fermionic unitary
circuits with the same complexity as for the corresponding spin model.
This opens up a way to efficiently describe fermionic higher-dimensional models without
intrinsic algorithmic problems, such as the sign problem in quantum Monte Carlo sampling. 
It is the hope that this work stimulates further 
interest in numerically assessing strongly correlated fermionic models
using unitary circuits.

{\it Note added:} We acknowledge discussions with U. Schollwoeck. We also
warmly acknowledge an early key discussion with F. Verstraete, and both this
work and Ref.\ \cite{VidalFermions} were sparked off from this initial discussion. In the
meantime, further approaches to the problem have appeared~\cite{Fermionsextra}.

{\it Acknowledgements.} This work was
supported by the EU (QAP, COMPAS, MINOS, QESSENCE), the EURYI, the 
CONACyT and the grant UNAM-PAPIIT IN117310.


\begin{thebibliography}{99}
\bibitem{footnote}
	 Its spin representation $s_{j,k}=J_{O^{t+1}}(S_{j,k})$ is given by  $s_{j,k}= |0,0\rangle\langle 0,0|
	+|0,1\rangle\langle 1,0| +|1,0\rangle\langle 0,1| -|1,1\rangle\langle 1,1|$.
\end{thebibliography}
\end{document}